\documentstyle[twocolumn,aps]{revtex}

\begin{document}

\twocolumn[\hsize\textwidth\columnwidth\hsize\csname 
               @twocolumnfalse\endcsname



\title{The Number of Particles Released in\\
Gauge Cosmic String Formation}


\author{Bj\o rn Jensen
\footnote{Electronic address: BJensen@boson.uio.no}}
\address{
Institute of Physics, University of Oslo,\\
P.O. Box 1048, N-0316 Blindern, Oslo 3, Norway}


\bibliographystyle{unsrt}

\maketitle

 
\begin{abstract}
We give a simple derivation of
 a new formal expression for the number of particles produced
from a conformal scalar field vacuum
due to the creation of a gauge cosmic string. 
We find that the number of particles
released in string formation
may be substantially lower than what previous
estimates have indicated. 
Our derivation also indicates that there
always exists at least one
critical angle deficit less than $2\pi$,
at which the particle production attains a maximum value.  
At the end we argue that additional
(quantum) effects will occur in string formation.
In particular, a new mechanism to produce small scale structure on
strings is proposed. {\em PACS nos.: 04.60\,\, ,\,\, 98.80.C}
\end{abstract}

\vskip2pc]




Gauge cosmic string theory is of great interest for the understanding
of both the origin, and the nature of the large
 scale structures observed in the universe.
Interest in cosmic string physics has recently increased also in other 
branches
of theoretical physics. In particular,
the connection between gauge cosmic string
physics, and the physics of black holes has  
been central to a number of recent remarkable advances in our
general understanding of the role played by black holes in nature,
on the character of black hole entropy, and the problem of
information loss in systems containing black holes \cite{Jensen1,Hawking1}. 

This paper deals primerely
with the response of the quantum vacuum due to the
creation of a gauge
cosmic string, but we will make contact with black hole
physics in order to facilitate our investigation.
We will in particular derive both a
quantitative estimate for the
number of particles
produced from the quantum vacuum during the formation
of such a string, as well as to seek a qualitative physical
understanding of the mechanisms involved in particle production in
string formation.
An extensive literature is already aviable on this 
important subject. In
this brief communication
 we are able to contribute to this fascinating field
by utilizing
a new expression for the geometries outside a static, as well
as a time-dependent, gauge cosmic string.
These geometries have been given a detailed
treatment elsewhere \cite{Jensen1,Jensen2}. They are briefly
reviewed below. We then derive
a new expression for the number of
produced particles from the quantum vacuum defined
by a scalar field, which is conformally coupled to gravity, when
a string is formed. We also relate our findings 
to a recent work on
a black hole-string system \cite{Jensen1} in order to enhance our
understanding, and extend the validity of our conclusions about
particle production in string formation.
At the end we argue that additional
(quantum) effects will occur in string formation which have escaped
the attention of previous works.

{\bf String Induced Geometries:} In order to deduce the geometric structure  outside of a gauge cosmic
string, we will assume that the string action can be approximated by
the action for a relativistic bosonic string \cite{Vilenkin1}. For our
purposes it is convenient to write this action as
\begin{eqnarray}
S&=&-\frac{\mu_0}{2}\int d^2\sigma\sqrt{-h}h^{\alpha\beta}g_{\mu\nu}
\partial_\alpha X^\mu\partial_\beta X^\nu\, .
\end{eqnarray}
$h_{\alpha\beta}=h_{\alpha\beta}(\tau ,\sigma )$
is the world sheet metric,
$X^\mu =X^\mu (\tau ,\sigma )$ the target space coordinates (i.e. the coordinates
describing the position of the string in space time),
$g_{\mu\nu}=g_{\mu\nu}(X^\mu )$ the target space geometry,
 and the pair $\tau$, $ \sigma$ represent the world-sheet
coordinates. We will follow standard practice and take
 $\tau$ to represent the intrinsic
time coordinate on the world sheet. The scalar $\mu_0$ represents the 
{\em constant} energy density
of the string as measured by a local observer, i.e. relative to
a local viel bein ``soldered'' to the string.  

A fundamental assumption in open string theory is that no point
on the string is different from any other except for the end points
of the string. This implies in particular that the world
sheet geometry can be made invariant under ``boost'' transformations. 
Hence, we may write $h_{\alpha\beta}=\Lambda\eta_{\alpha\beta}$, where
$\eta_{\alpha\beta}$ is the two dimensional Minkowski metric,
and $\Lambda$ is an arbitrary scalar function of the world sheet coordinates,
and possibly the constant factor $\mu_0$. We will be concerned with a straight,
infinitely long, and 
to begin with, a time independent string.
This means in particular that it is natural to orient the string 
in such a way that $\tau$ coincides
with the target space time coordinate $t$, and $\sigma$ with the
spatial target space coordinate along the string $z$. In this picture it is
furthermore natural to assume that $h_{\alpha\beta}$ is the geometry
induced on the world sheet by the space time metric $g_{\mu\nu}$.
To conform with the
picture of an infinitely long and straight string, we demand manifest
Poincare' invariance on the world sheet, which further restricts $\Lambda$ to
equal a coordinate independent function
$\Lambda_0=\Lambda_0(\mu_0 )$.
 We may now formally
use the conformal invariance of the string action
in order to bring it to a standard form.
Since $h_{\alpha\beta}=\Lambda_0\eta_{\alpha\beta}$
we have $h=-\Lambda_0^2$, $h^{\alpha\beta}=\Lambda_0^{-1}\eta^{\alpha\beta}$,
which means that the action can be written as
$S=-\frac{\mu_0}{2}\int d^2\sigma\eta^{\alpha\beta}g_{\mu\nu}
\partial_\alpha X^\mu\partial_\beta X^\nu$.
However, even though it seems from this last ``exercise'' that
$\Lambda_0$  always
can be transformed away,
this is true modulus some key assumptions. One must in particular assume that
this factor is {\em non singular} in order to have a well defined transformation. 
In our approach we will see that this is not always the case.

From the action $S$ we can deduce the energy momentum tensor for a relativistic
string. In the limit when the core radius of the string is set equal
to zero, it can be shown \cite{Jensen1,Jensen2} that the resulting induced
geometry outside a static string
is given by
\begin{eqnarray}
ds^2&=&\frac{1}{(1-4\mu_0)}(-dt^2+dr^2+dz^2+r^2(1-4\mu_0)^2 d\phi^2)\\
&\equiv& g_{\mu\nu}dx^\mu dx^\nu\equiv D_0^{-2}G_{\mu\nu}dx^\mu dx^\nu\, .
\end{eqnarray}
Here we have defined $D_0^2\equiv (1-4\mu_0 )$ for 
later convenience. 
When we employ the globally defined
$(t,r,z,\phi )$ coordinate system, it follows that we may
identify $D_0$ with $\Lambda_0$, i.e. we may set $\Lambda_0=D_0^{-2}$.
Note that $\Lambda_0$ becomes singular in the limit $\mu_0\rightarrow\frac{1}{4}$.
Previous
derivations of the geometry outside a string
have failed to recognize the presence of the ``conformal''
factor $D_0^{-2}$ in the exterior string geometry.
The presence of this ``conformal'' factor
is vital for a proper understanding of the nature of
the gravitational field outside a string \cite{Jensen1,Jensen2}.
It will also be of significant importance when
we consider quantum processes outside
such a string during its formation. Eq.(2)
can be brought to the standard form for the string induced geometry
by the coordinate transformation $t\rightarrow D_0t$, $r\rightarrow
 D_0r$, $z\rightarrow D_0 z$ provided that
$\mu_0$ is a constant factor and $\mu_0<\frac{1}{4}$.
Hence, in the static situation the ``conformal'' factor will not
give rise to any new physical effects.  
Previous calculations of the energy production
due to the creation of a string employed the $G_{\mu\nu}$ geometry
with $\mu_0\rightarrow \mu =\mu (t)$ as an approximate description of the
real string formation process. 
It is clear that $g_{\mu\nu}$ cannot be
transformed into $G_{\mu\nu}$ when $\mu_0\rightarrow \mu =\mu (t)$ 
neither with the
help of the coordinate transformation above nor 
with the help of any other such
transformation. In the time dependent situation the ``conformal''
factor will thus be a ``true'' conformal factor (when we assume that
$\mu_0<\frac{1}{4}$) and not ``merely'' a factor which can be
transformed away with the help of a coordinate transformation.
Hence, new quantum effects will in general be expected to appear
in the string creation process 
when eq.(2) is used as the background geometry with $\mu_0\rightarrow
\mu =\mu (t)$ when
compared with previous such studies
which employ the $G_{\mu\nu}$ geometry.

As an aside it is of some interest to ask whether it is possible to
demonstrate that the geometry eq.(2) can be reached from,
say, Minkowski space time. In \cite{Jensen2} this was partially achieved
when the metric ansatz was taken on the form
$ds^2=e^{2\Psi}(-dt^2+dr^2+dz^2+A^2e^{-4\Psi}d\phi^2)$.
Here it is assumed that both
$\Psi$ and $A$ only depend on the time like and radial
coordinates. Einstein's field equations
{\em without any other sources than a straight (singular) string} can be 
partially integrated in a straight forward
fashion. One note worthy solution for $\Psi$ is on the form
$e^{2\Psi}\sim (1-C_1e^t)^4e^{-2t}e^{-2r}$ ($t\geq 0\, ,\, r\geq 0$)
where $C_1$ is an integration constant.
For sufficiently large times we have $e^{2\Psi}\sim
e^{2t}e^{-2r}$, i.e. $|g_{tt}|$ is a {\em monotonically increasing}
function of time
\footnote{Interestingly, a similar
de Sitter like behavior was also found in a recent numerical study of the
creation of a {\em global} cosmic string {\em gr-qc/9606002}.}.
 This behavior is in correspondence
with the form of the geometry eq.(2),
 since $|g_{tt}|\geq 1$ for a static string. However, $e^{2\Psi}$ 
seems to grow with {\em no upper bound}. This is 
a signal that our model probably
is to simple, and that a more realistic model
should also take into account the underlying field theoretical
model in which the string actually lives. This will probably induce
a non vanishing $T_{\phi\phi}$ component
in the energy momentum tensor of the more complete system. This 
may modify the time evolution of the string geometry
in this regime. However, we believe that we have grasped some of the
qualitative behavior of this field, and that we have shown ``beyond
reasonable doubt'' that the metric outside a static string can be taken on the
form of eq.(2), whenever the action $S$ is a good approximation to the
full string effective action.

{\bf Particle production in string formation :} To obtain 
an expression for the number of particles produced from the
quantum vacuum due to the
creation of a gauge cosmic string, we will follow the possibly
simplest approach, the so called
sudden approximation \cite{Parker}.
Although this approach has been criticized for not, in
some sense, being ``analytical'', i.e. it relies necessarily
on non differentiable mode solutions, it is in general
believed to give the correct order of magnitude for the 
number of produced particles \cite{Mendel}.
In the sudden approximation we assume
space time to be flat Minkowski space for $t<t_1$, while the
space time geometry is given by eq.(2) when $t>t_1$ for some
appropriately chosen global time $t=t_1$, i.e. we follow
previous studies and write $\mu (t)=
\mu_0\theta (t-t_1)$ where $\theta$ is the step function.
Previous estimates for the number of
particles released during the formation of
a string, are based on the geometry $G_{\mu\nu}$. Our
strategy is to establish a formal relation between the 
number of particles produced
$|\beta_G|^2$ relative to $G_{\mu\nu}$,
and the corresponding expression $|\beta_g|^2$ relative to $g_{\mu\nu}$. 
 
We will first concentrate on the simplest 
(in our case) scalar field configuration, namely
that of a conformally coupled and mass less scalar field, with the action
\begin{equation}
S=\int d^4 x\sqrt{-g}(\partial_\mu\Phi\partial^\mu\Phi 
-\frac{1}{6}R\Phi^2)\, .
\end{equation}
The reason for choosing this action is that 
the associated equation of motion for $\Phi$ is ``invariant''
under space time conformal transformations. We will focus on the conformal
relation $g_{\mu\nu}=D_0^{-2}G_{\mu\nu}\equiv\Omega^2G_{\mu\nu}$
given in eq.(3). Since the
scalar field is assumed to be conformally coupled to gravity,
the equations of motion of the scalar field $\Phi_g$,
relative to $g_{\mu\nu}$, and the corresponding equation for
the scalar field relative to $G_{\mu\nu}$ are related by
\begin{equation}
(\Box_g+\frac{1}{6}R_g)\Phi_g=\Omega^{-3}(\Box_G+\frac{1}{6}R_G)\Phi_G=0\, ,
\end{equation}
where the following relation between the fields holds
$\Phi_g=\Omega^{-1}\Phi_G$ \cite{Birrell}. The operator
$\Box$ is the d'Alembertian relative to the geometry indicated.
We second quantize the scalar field $\Phi_G$ 
in the asymptotic past
such that this field
is promoted to an operator $\hat{\Phi}_G$, which we expand as
\begin{equation}
\hat{\Phi}_G=\sum_{n}(\hat{a}^{\mbox{in}}_{Gn}\Phi^{\mbox{in}}_{Gn}+
\hat{a}^{\mbox{in}\dagger}_{Gn}\Phi^{\mbox{in}*}_{Gn})\, ,
\end{equation}
in terms of a complete set of solutions $\{\Phi^{\mbox{in}}_{Gn}\}$ of the
equation of motion. 
$n$ indicates the collection of quantum numbers needed to
characterize these solutions.
Likewise, in the asymptotic future we expand the
fields in terms of a complete set of mode solutions 
$\{\Phi_{Gn}^{\mbox{out}}\}$ and operators $\hat{a}^{\mbox{out}}_{Gn}$.
Relative to the in set of operators we may define a natural vacuum state
by $\hat{a}^{\mbox{in}}_{Gn}|0;G,\mbox{in}\rangle =0$,
while we relative to the out set in general
will get {\em another} vacuum by $\hat{a}^{\mbox{in}}_{Gn}|0;G,\mbox{out}
\rangle =0$. The vacua are different
in the sense that $|0;G,\mbox{out}\rangle$ in general
will be a mixed state when expressed
in terms of the Hilbert space defined by the in vacuum state.
This mixing can be expressed by
\begin{equation}
\Phi_{Gn}^{\mbox{out}}=\sum_{m}(\alpha_{Gmn}\Phi_{Gm}^{\mbox{in}}+
\beta_{Gmn}\Phi_{Gm}^{\mbox{in}*})\, ,
\end{equation}
where $\alpha_{Gmn}$ and $\beta_{Gmn}$ are known coefficients.
On the assumption that the field is in the vacuum state in the asymptotic
past, the total number of particles produced during the formation of a string
can formally be written as
\begin{eqnarray}
|\beta_G|^2&=&\sum_{nn'}\langle\mbox{in},G;0|
\hat{N}^{\mbox{out}}_{Gnn'}|0;G,\mbox{in}\rangle\, ,
\end{eqnarray}
where $\hat{N}^{\mbox{out}}_{Gnn'}=\hat{a}^{\mbox{out}}_{Gn}\hat{a}^{\mbox{out}
\dagger}_{Gn'}$. A complete set of mode functions in the in region 
is in a rather self explanatory form given by
(with a slight abuse of notation)
\begin{equation}
\Phi_{Gn}^{\mbox{in}}=\frac{e^{-i\omega_n^{\mbox{in}}t}}{\sqrt{2\pi\omega_n^{\mbox{in}}}}
e^{ik_n^{\mbox{in}}z}e^{im\phi}J_{|m|}(k^{\mbox{in}}_nr)\, ,
\end{equation}
where $m$ is an integer,
with respect to the canonical inner product $(\, ,\, )_G$ defined by 
\begin{equation}
(\Phi_n,\Phi_{n'})_G=i\int_\Sigma d^3x\sqrt{-G}\Phi_n{\cal
 L}_{\vec{\xi}}\Phi_{n'}^*\, .
\end{equation}
The integral is in general taken over some appropriately defined
Cauchy hyper surface $\Sigma$. 
${\cal L}_{\vec{\xi}}$ denotes the Lie derivative in some
time like direction. We will choose $\vec{\xi}$ to coincide with the
canonical time direction in Minkowski space time.
Hence, this derivative defines our 
usual notion of positive and negative
frequencies.
Even though we will assume a {\em singular} string source,
we will work as if space time is globally hyperbolic, such that this product always
makes sense, and is thus conserved along time like trajectories. The corresponding
set of modes in the out region is
\begin{equation}
\Phi_{Gn}^{\mbox{out}}=\frac{e^{-i\omega_n^{\mbox{out}}t}}{\sqrt{2\pi
\omega_n^{\mbox{out}}}}
e^{ik_n^{\mbox{out}}z}e^{im\phi}
J_{\alpha |m|}(k^{\mbox{out}}_nr)\, ,
\end{equation}
where $\alpha^{-1}=(1-4\mu_0)$.
It has been shown in a large body of previous work,
that the two sets can be
connected in a non trivial way such that $\omega^{\mbox{out}}\, ,
\, k^{\mbox{out}}$ in
general will be non trivial and analytical functions of $\omega^{\mbox{in}}
\, ,\, k^{\mbox{in}}$.
This leads to the well known result that particles, as this concept
is defined relative to $|0;G,\mbox{in}\rangle$, are present in the
canonical out vacuum state $|0;G,\mbox{out}\rangle$, i.e. $|\beta_G|^2\neq 0$.

How is this picture changed when we consider this discussion using the
$g_{\mu\nu}$ geometry ? In the asymptotic past, space time is flat
Minkowski space time, and we may therefore identify $\Phi^{\mbox{in}}_{gn}=
\Phi^{\mbox{in}}_{Gn}$, $\hat{a}^{\mbox{in}}_{gn}=\hat{a}^{\mbox{in}}_{Gn}$,
and hence set $|0;g,\mbox{in}\rangle =|0;G,\mbox{in}\rangle$. Since we are
dealing with a scalar field which is conformally
coupled to gravity we know that $\Phi^{\mbox{out}}_{gn}=\Omega^{-1}\Phi^{\mbox{out}}_{Gn}$.
The new vacuum state which arise when this relation is put
into a similar operator relation
is a so called conformal vacuum state \cite{Birrell}.
When this last relation is applied to eq.(7) we find
\begin{equation}
\Phi^{\mbox{out}}_{Gn}=\sum_{m}(\Omega\alpha_{gmn}\Phi^{\mbox{in}}_{gm}
+\Omega\beta_{gmn}\Phi^{\mbox{in}}_{gm})\, .
\end{equation}
Hence we have that
$\alpha_{gmn}=\Omega^{-1}\alpha_{Gmn}$, $\beta_{gmn}= \Omega^{-1}\beta_{Gmn}$.
It follows that 
$\hat{N}_{gnn'}^{\mbox{out}}=\Omega^{-2}\hat{N}_{Gnn'}^{\mbox{out}}$. We are
then lead to the following expression for the total number of produced
particles relative to the conformally related geometry in the
sudden approximation
\begin{eqnarray}
|\beta_g|^2=\Omega^{-2}|\beta_G|^2=(1-4\mu_0)|\beta_G|^2\, .
\end{eqnarray}
We may view this expression for the production 
of particles in the conformally related geometry
as being composed out of two contributions. One 
contribution comes from the existence of
the angle deficit that the string cuts out, and which is coded in $|\beta_G|^2$.
The other contribution stem from the conformal factor which makes up the rest
of the expression for $|\beta_g|^2$. How is this part
of the expression to be understood ? 

The origin of $|\beta_G|^2$ can be seen
as due to the presence of a ``Casimir'' like
energy in the vacuum in the $G$ geometry. 
Since a string cuts out an angle deficit, a negative energy
density is induced in the quantum vacuum defined by the operator
$\hat{\Phi}_G=\hat{\Phi}_G(\Phi_{Gn}^{\mbox{out}})$. Due to 
overall energy
conservation, positive energy must have been removed from the
initial Minkowski vacuum. This positive energy is realized as 
real particles,
and is reflected in a non vanishing $|\beta_G|^2$. The conformal
factor in eq.(2) ``stretches'' all proper lengths, i.e. the volume
defined to be inside a box with the walls at fixed coordinate positions
in the $G$ geometry, is expanded. When it comes
to vacuum fluctuations this implies a red shift effect. Hence,
due to this effect the ``Casimir'' energy will be smaller in
absolute value than the corresponding energy computed without the
conformal factor present. This can be seen by a direct computation of the
energy momentum tensor of the scalar field. Using the point splitting method
\cite{Birrell}, the central function from which this tensor is computed is the
two point correlation function $G(x,x')$. In general we will have
$G_G(x,x')\rightarrow G_g(x,x')=\Omega^{-1}(x)G_G(x,x')\Omega^{-1}(x')=
(1-4\mu_0)G_G(x,x')$. This implies a similar relation between the
corresponding energy momentum tensors. This verifies our
qualitative picture. This simple picture can be further
substantiated with a comparison with the
black hole-string composite system which was studied in \cite{Jensen1}.
There it was shown
that the Hawking temperature $T$ was {\em lower} for a black hole with a string
through it, than the temperature of the corresponding black hole
without a string. Since we have been studying the production
of mass less scalar particles we may approximate the relation between
the mean number of particles $\overline{N}$
produced by the black hole, and its temperature $T$, by 
$\overline{N}\sim T^3$. This relation
holds exactly for a photon gas in flat space.  
Our present analysis is in line with this relation,
since a lower temperature is correlated with a decrease of the mean
number of particles. 

This connection to black hole physics seems to imply
that the ``stretching'' effect has the tendency to lower the
production of particles in any particle creating process. 
Since the black hole temperature is
universal in the sense that all particle species will be radiated
away by the black hole at {\em the same temperature}, it is to be
expected
that the reduced production estimate we have derived is probably not
particular to just the conformal 
scalar field sector. We expect that the production
estimates for all particle species in the process of 
string formation from flat space, will be lower compared to previous
calculations. In cosmic string formation a picture of two 
competing processes
emerges. The appearance of a conic angle deficit leads to
particle production, while the conformal factor has the tendency to
lower the production compared with the situation when this factor
is not present.
 For small $\mu_0$ we expect that
$|\beta_G|^2\sim\mu_0^2$ \cite{Parker,Mendel}.
 Hence, in this regime $|\beta_g|^2$ is
a growing function of $\mu_0$. However, $|\beta_g|^2$ can be made,
at least formally,
as close
to zero as one wishes by letting $\mu_0\rightarrow\frac{1}{4}$. 
It follows that this function must have at least one maximum point
between $\mu_0=0$ and $\mu_0=\frac{1}{4}$. Hence, for sufficiently
large values of $\mu_0$ the ``stretch'' effect 
will dominate, such that
a further increase of the value in this parameter does not lead to an
increased particle production. However, our study does not indicate
whether the maximum point is located at relatively low energies or
at higher energies (i.e. $\mu_0\sim\frac{1}{4}$). The location of this
point may also depend on the particular quantum system under
consideration. It is therefore difficult to derive an
explicit numerical estimate for the energy production from string
formation, but is clear from our considerations that
the deviation from earlier studies may be substantial. However,
due to the general complexity of the equations of motion for 
fields living in the time dependent ``version'' of eq.(2)
when the time evolution of the string mass
parameter is taken to be other than the step function, such
estimates can probably only be obtained by numerical means. 

We have not exhausted all possibilities for
interesting quantum effects during the formation of a cosmic string.
It is tempting to take the analogy with the de Sitter space,
which we noted earlier, seriously. In that space particle horizons exist,
which are closely related to event horizons also on the quantum level.
That a horizon structure in fact is present in our case is easy to
see since the proper linear velocity of any fixed point 
(relative to the $(t,r,z,\phi )$ coordinates)
on the string, or in the radial direction,
relative to any other fixed point on the string
is given by $v=HD$. The ``Hubble factor'' $H$ is given by
$H=\Omega^{-3}\dot{\Omega}$, and $D$ is the proper distance between
the points. Clearly an horizon is present when $v=1$, i.e. at
$D=H^{-1}$. Due to this horizon structure  
particles will be expected to be produced in addition to those already
considered above. The same fluctuations which produce these
particles are also expected (in analogy with de Sitter space)
to back react on the string, and
thus give rise to strings which are no longer straight \cite{Hawking2}.
This might represent a viable mechanism to produce strings
which can fit the description of ``wiggly'' strings \cite{Vilenkin1},
i.e. strings with small scale structure.
These possible new effects are missed in this work due to the nature of
the sudden approximation. It would be very interesting,
although probably
very difficult due to a time and position dependent ``Hubble
factor'', to examine these questions further in greater detail.


\begin{thebibliography}{999}

\bibitem{Jensen1} B. Jensen, Nucl.Phys.B {\bf 453}, 413 (1995).

\bibitem{Hawking1} S.W. Hawking and S.F. Ross, Phys.Rev.Lett. {\bf 75},
3382 (1995); R. Emparan, {\em ibid.} {\bf 75}, 3386 (1995); 
D.M. Eardley, G.T. Horowitz,
D.A. Kastor and J. Traschen, {\em ibid.} {\bf 75}, 3390 (1995);
 R. Gregory and M. Hindmarsh,
Phys.Rev.D {\bf 52}, 5598 (1995); A. Achucarro, R. Gregory and
K. Kuijken, {\em ibid.} {\bf 52}, 5729 (1995);
R. Emparan, Phys.Rev.D {\bf 52}, 6976 (1995); F. Englert, L. Houart
and P. Windey, Phys.Lett.B {\bf 372}, 29 (1996); Nucl.Phys.B {\bf 458}
231 (1996).

\bibitem{Jensen2} B. Jensen, {\em Geometries of Relativistic Strings
and other $p$-branes},
Oslo-TP 2-96, J. Math.Phys. (in press). 

\bibitem{Vilenkin1} A. Vilenkin and E.P.S. Shellard, {\em
Cosmic strings and other topological defects}, Cambridge UP (1994). 

\bibitem{Parker} L. Parker, Phys.Rev.Lett. {\bf 59}, 1369 (1987). 

\bibitem{Mendel} G. Mendel and W.A. Hiscock, 
Phys.Rev.D {\bf 40}, 282 (1989). 

\bibitem{Birrell} N.D. Birrell and P.C.W. Davies, {\em Quantum fields
in curved space}, Cambridge UP (1986).  

\bibitem{Hawking2} S.W. Hawking, 
Phys.Lett.B {\bf 115}, 295 (1982).

\end{thebibliography}
\end{document}